\documentstyle[aps,prl,preprint,floats,epsfig]{revtex}

\begin{document}

\preprint{\tighten\vbox{\hbox{\hfil CLNS 01/1769}
                        \hbox{\hfil CLEO 01-24}
}}

%\title{Your Title Goes Here}  
\title{Further Experimental Studies of Two-Body Radiative $\Upsilon$ Decays}

% Your author list ***DOES NOT*** go here!
% is goes below where you are instructed to insert it...
\author{CLEO Collaboration}
% You will want to hard code the date once you are ready to submit your paper!
\date{\today}

\maketitle
\tighten

\begin{abstract} 
% Insert abstract here.
Continuing our studies of radiative $\Upsilon$ decays,
we report on a search for 
$\Upsilon\to\gamma\eta$ and
$\Upsilon\to\gamma f_{J}$(2220) in 
61.3 pb$^{-1}$ of 
$e^{+}e^{-}$
data taken with the CLEO II
detector at the Cornell Electron Storage Ring.  
For the $\gamma\eta$ search the three decays of the $\eta$ meson
to 
$\pi^{+}\pi^{-}\pi^{0}$, $\pi^{0}\pi^{0}\pi^{0}$,
and $\gamma\gamma$ were investigated.
We found no candidate events in the $(3\pi)^{0}$ modes and
no significant excess over expected backgrounds in the $\gamma\gamma$
mode to set a limit on the branching fraction of 
${\cal B}(\Upsilon\to\gamma\eta)$$<$$2.1\times 10^{-5}$ 
at 90\% C.L.
The three
charged two-body final states 
$h\overline{h}$ ($h = \pi^{+}, K^{+}, p$)
were investigated for $f_{J}$(2220) production, with 
one, one, and two events found, respectively.  Limits at  90\% C.L.
of
${\cal B}(\Upsilon\to\gamma f_{J})\times{\cal B}(f_{J}\to h\overline{h})$
$\sim 1.5 \times 10^{-5}$
have been set for each of these modes.
We compare our results to 
measurements of 
other radiative $\Upsilon$   decays, to 
measurements of 
radiative $J/\psi$ decays, and 
to theoretical predictions.

\end{abstract}
\newpage

{
\renewcommand{\thefootnote}{\fnsymbol{footnote}}

% Insert author and address list here

\begin{center}
G.~Masek,$^{1}$ H.~P.~Paar,$^{1}$
R.~Mahapatra,$^{2}$
R.~A.~Briere,$^{3}$ G.~P.~Chen,$^{3}$ T.~Ferguson,$^{3}$
G.~Tatishvili,$^{3}$ H.~Vogel,$^{3}$
N.~E.~Adam,$^{4}$ J.~P.~Alexander,$^{4}$
% ...
\end{center}
 
\small
\begin{center}
$^{1}${University of California, San Diego, La Jolla, California 92093}\\
$^{2}${University of California, Santa Barbara, California 93106}\\
$^{3}${Carnegie Mellon University, Pittsburgh, Pennsylvania 15213}\\
$^{4}${Cornell University, Ithaca, New York 14853}\\
% ...
\end{center}

\setcounter{footnote}{0}
}
\newpage

% Insert body of the text here.

\section{Motivation}

Exclusive radiative decays of 
the heavy vector states $J/\psi$ and
$\Upsilon$ have been the subject of many
experimental and theoretical studies.  For the 
experimenter, the final states from $V \to\gamma R$
are easy to identify and measure in that they
have a high energy photon and low multiplicity
of other particles.  Backgrounds also tend to be
small.  Theoretically the emission of the photon
leaves behind a glue-rich environment from
which to learn about the formation of resonances
from two gluons or 
to
discover new forms of hadronic
matter.
Because gluons ($g$) themselves carry the quantum number
of ``color'', QCD %the theory 
allows for 
states
with {\it no} valence quarks ($q$), but only gluon
constituents: ``glueballs''.  QCD
also allows for more exotic combinations such as $qg\overline{q}$
``hybrids''.  These glueballs and hybrids are not just more
resonances in the spectrum - they represent fundamentally different
forms of matter from the mesons and baryons with which we are so
familiar.

For the $J/\psi$ charmonium system (the  $1_{3}S^{1}$
state of $c\overline{c}$) many such radiative two-body decays
have been observed\cite{PDG2000}, with some of the
dominant ones being $\gamma f_{2}$(1270), $\gamma\eta$,
and $\gamma\eta^{\prime}$.  However, the only such
observation in the radiative decay of the $\Upsilon$
(the  $1_{3}S^{1}$ state of $b\overline{b}$)
is by CLEO\cite{Korolkov} in the final state
$\gamma\pi\pi$, in which an enhancement
in the di-pion invariant mass consistent with
being the $f_{2}$(1270) meson
was observed.  A recent CLEO search\cite{etaprime} for the
radiative
production of the $\eta^{\prime}$ meson
yielded only an upper limit.

In this
Article
we present a search for the radiative production
of the other isoscalar pseudoscalar, namely the $\eta$ meson.
As with the final state $\gamma\eta^{\prime}$, this channel 
has received significant theoretical attention. %dzb in the literature. 
The work of K\"orner, K\"uhn, Krammer, and Schneider~\cite{KKKS}
and the followup publication by  K\"uhn %dzb: alone~
\cite{Kuhn83} use
highly virtual gluons to predict minimal suppression of radiative
pseudoscalar production as the vector meson mass goes from that
of the $J/\psi$ to that of the $\Upsilon$.  
Intemann~\cite{Intemann} used the vector dominance model (VDM) %dzb-change
to predict branching fractions for $\Upsilon\to\gamma\eta$ 
taking into account the
interference between the $\Upsilon$ and $\Upsilon^{\prime}$,
the major contributing vector mesons in the model.
Baier and Grozin \cite{Baier} showed that for lighter
vector mesons (such as the $J/\psi$) there might be an additional 
``anomaly'' diagram that contributes significantly to the radiative decays.
Ball, Fr\`ere, and Tytgat
worked along similar lines~\cite{Ball}. However, Baier and Grozin note that
their approach applies directly to the ``singlet'' member of the
meson nonet.  Feldmann, Kroll, and Stech\cite{FKS}
pursue the ideas of mixing in the decay constants of the
pseudoscalars to derive ratios of their radiative production.
Chao\cite{Chao} has taken this approach further,
determining mixing angles such as
$\lambda_{\eta\eta_{b}}$ between the $\eta_{b}$ and $\eta$
in order to calculate radiative branching fractions.  
Finally, the recent work of Ma Jian-Ping~\cite{Ma} uses factorization 
at tree
level with
non-relativistic QCD matrix elements 
to describe the heavy vector meson
portion multiplied by a set of twist-2 and twist-3 gluonic
distribution amplitudes.  

The other search we present here is for the 
radiative production of the $f_{J}$(2220),
also known as the $\xi$(2230), in $\Upsilon$ decay.
Many theoretical 
calculations
of the spectrum of glueballs
predict a $J^{PC} = 2^{++}$ state
in the area of 2.2 GeV/$c^{2}$.  A candidate for this tensor
glueball has been seen by some experiments, but not by 
others\cite{PDG2000}.  The most complete claim of observation 
is by the BES collaboration who have published results\cite{BES}
for $J/\psi \to \gamma f_{J}$(2220) with the $f_{J}$(2220) reconstructed
in $\pi^{+}\pi^{-}$, $K^{+}K^{-}$, $p\overline{p}$,
and $K^{0}_{S}K^{0}_{S}$ as well as for
$\pi^{0}\pi^{0}$\cite{BESpi0}
and for
$\eta\eta^{\prime}$\cite{BESconf}.
Of significant interest are several
non-observations.  CLEO has not seen the $f_{J}$
in two-photon interactions\cite{CLEOgamgam}, which
would lend credence to its being a glueball.
However, 
a narrow resonance in this mass
region was not seen 
in $p\overline{p}$ production by either JETSET
\cite{JETSET} 
or, more recently, by Crystal Barrel\cite{XBarrel}.
This non-observation sheds doubt on the
very existence of the $f_{J}$.

\section{Detector and Data Sample}

Our analyses used 61.3 pb$^{-1}$ of 
$e^{+}e^{-}$
data recorded at the
$\Upsilon$(1S) resonance ($\sqrt{s} = 9.46$ GeV) with the
CLEO II detector\cite{CLEOII} 
operating at the Cornell Electron Storage Ring (CESR).
This corresponds to the production of 
$N_{\Upsilon} = (1.45 \pm 0.03)\times 10^{6}$
$\Upsilon$(1S) mesons\cite{Korolkov}.
In addition, 
significantly larger samples taken near in time to this
$\Upsilon$(1S) data but at energies at or just below the $\Upsilon$(4S)
were used for comparison to the underlying continuum.
The momenta and ionization loss ($dE/dx$) 
of charged tracks were measured in a six-layer straw-tube chamber, 
a ten-layer 
precision drift chamber, and a 51-layer main drift chamber, all
operating in a 1.5~T solenoidal magnetic field. 
Photon detection and electron suppression were
accomplished using the high-resolution electromagnetic 
calorimeter consisting of 7800 thallium-doped CsI crystals.
The work presented here used only events
with the primary, high-energy photon %dzb??? use 
in the barrel portion of this detector,
defined as $|\cos{\theta}| \leq 0.71$, 
because the energy resolution 
for photons and reconstruction
efficiency of the recoiling neutral
mesons are degraded in the 
end cap regions and because
the efficient, well understood
triggers involve only the barrel region of the
calorimeter. 
Between the central drift chamber and the electromagnetic
calorimeter, strips of scintillating plastic were used for
triggering and for measuring time-of-flight.
Proportional tracking chambers for muon identification
were located between and outside the iron slabs that provide
the magnetic flux return.
The Monte Carlo simulation 
of the detector response was based upon GEANT\cite{GEANT}, and simulation
events were processed in an identical fashion to data.

\section{Search for $\Upsilon\to\gamma\eta$}

Our search for 
$\Upsilon \to \gamma \eta$ involved the decays 
$\eta\to\gamma\gamma$,
$\eta\to\pi^{0}\pi^{0}\pi^{0}$, or 
$\eta\to\pi^{+}\pi^{-}\pi^{0}$;
the latter two will collectively be referred
to as $(3\pi)^{0}$.
We followed procedures very similar to those used
in our recent publication on the $\gamma\eta^{\prime}$
final state\cite{etaprime}.
In order
to maximize detection efficiency and minimize possible
systematic biases, we employed a minimal number of selection criteria,
with combinatoric background largely
suppressed by requiring reconstruction of both the
$\Upsilon$ and $\eta$ mesons.

Events were required to have the proper number of quality tracks 
(either zero or two) of 
appropriate
charges
and at least three calorimeter 
energy clusters
(which may or may not
be associated with the tracks), of which one had to correspond to an energy
of at least 4 GeV and be in the barrel fiducial volume
($|\cos\theta|$ $\leq$ 0.71).  In addition, 
we required that the events pass
trigger 
criteria
\cite{CLEOTrig}, 
based purely on the calorimeter, 
that were highly efficient and 
could be reliably simulated.

For reconstructing $\pi^{0}$ candidates, the photon
candidates had to have minimum
depositions of 30 (50) MeV in the 
barrel (endcap) 
regions\footnote{The endcap region 
is defined as
$0.85$$<$$ |\cos\theta| $$<$$0.95$; the region between this and the
barrel fiducial region is not used due to its poor resolution.} 
and could not be associated
with any charged track; in addition, at least one of the two photons
had to be in the barrel region. The $\gamma\gamma$ invariant mass 
had to be within
50 MeV/$c^{2}$ ($\sim \pm 9 \sigma_{\pi}$) of the known $\pi^{0}$ 
mass\cite{PDG2000}; such
candidates were then kinematically constrained to that mass.
The 
photon candidates
used in reconstructing the $\eta$ meson in $\gamma\gamma$
had
to deposit
a minimum of 60 (100)~MeV in the barrel (endcap) calorimeter regions,
could not be identified as a fragment of a charged track deposition,
and had to have a lateral profile consistent with that of a photon.

For the $\gamma (3\pi )^{0}$ modes we then built
$\eta$ candidates from
$\pi^{0}\pi^{0}\pi^{0}$ or $\pi^{+}\pi^{-}\pi^{0}$.  Simulation
events were used to determine the detector mass resolution for 
these two signal modes: $\sigma_{\eta} = 10.7$ and 9.0 MeV/$c^{2}$, respectively.
Candidates had to be within $\pm 3 \sigma_{\eta}$ of the known $\eta$ mass.
In the case of the $\gamma\pi^{0}\pi^{0}\pi^{0}$ final state, no photon
could be common to more than one $\pi^{0}$ 
combination.
To suppress QED backgrounds in the $\gamma\pi^{+}\pi^{-}\pi^{0}$
final state, a charged track was
rejected if its momentum, $p$, from the drift chamber matched
its energy, $E$, as measured in the calorimeter as
0.85 $<$$E/p$$<$~1.05.

Then, $\Upsilon$ candidates were formed by combining the high-energy
photon ($E$$>$4 GeV) with the $\eta$ candidate, 
requiring
that this photon
not already 
be
used in reconstructing the event.  To be considered,
such a candidate had to have an invariant mass within $\pm 300$ MeV/$c^{2}$
of $\sqrt{s} = m_{\Upsilon}$, 
a window of
roughly three times the detector
resolution as obtained from our simulations.
Although,  in general, multiple candidates per event were not restricted,
there were two exceptions:
(i) in the case of $\eta\to\pi^{0}\pi^{0}\pi^{0}$,
if two $\eta$ candidates shared more than four photons, the candidate
with the better combined $\chi^{2}$ for mass fits to 
the three $\pi^{0}$ candidates was accepted; and (ii) in the 
case of $\eta\to\pi^{+}\pi^{-}\pi^{0}$, if two candidates for the neutral
pion shared a daughter photon, the one with the better fit to the
$\pi^{0}$ mass was taken.

After these 
highly-efficient
procedures were applied, we found {\it no} candidates in either
the 61.3 pb$^{-1}$ of $\Upsilon$(1S) data  or 
in 189 pb$^{-1}$ of continuum data 
samples.\footnote{While some of these data were at the $\Upsilon$(4S)
resonance, we note that $B$ meson decays cannot have a high
energy photon so that 
the analysis of these data yields only 
$udsc$ or QED
backgrounds.}

From Monte Carlo simulations, the overall
efficiencies for each channel, $\epsilon_{i}$, were determined to
%Forgot correction for cos^2 theta and trigger; matches table now
%Also forgot to include effects of systematic uncertainty
%be $(8.5 \pm 0.2)\%$ and $(30.9 \pm 0.3)\%$
be $(7.6 \pm 0.8)\%$ and $(26.7 \pm 1.5)\%$
for the decay chains ending in 
$\eta\to\pi^{0}\pi^{0}\pi^{0}$ and 
$\eta\to\pi^{+}\pi^{-}\pi^{0}$, respectively.  The
uncertainties here include the statistics of the Monte
Carlo samples and our estimates on possible systematic biases,
which we discuss below.  
Including the branching fractions
for the $\eta$ decays\cite{PDG2000}
and their uncertainties gave
$\sum [\epsilon_{i}{\cal B}_{\eta,i}] = (8.7 \pm 0.5)\%$
for 
$\Upsilon\to\gamma(3\pi )^{0}$.

For the final state $\gamma\gamma\gamma$, the significant
QED background compelled us to change the order of the constraints
on the meson reconstruction and
to add one additional selection criterion.  
Here we {\it first} took three photons,
as defined above, and required that 
$|m_{\gamma\gamma\gamma} - m_{\Upsilon}| \leq$ 300 MeV/$c^{2}$.  Then
we took the photons in pairs and plotted the spectrum
of $m_{\gamma\gamma}$.

A large background exists in both the $\Upsilon$(1S) and
$\Upsilon$(4S) data sets, peaking near 0.40 GeV/$c^{2}$.  From scanning
such events it is evident that these are $e^{+}e^{-}\to\gamma\gamma$
events with pair production by one of the photons giving a final
state of $\gamma e^{+}e^{-}$.  Tracks were not reconstructed in this subset of
events due to the timing characteristics of the energy-based triggers
in CLEO~II.  
Such conversion events have their showers separated
only in azimuth, in that the lepton pair has zero opening angle and any
observed separation is due only to the bending by the magnetic field.
Therefore, to
suppress this
$\gamma e^{+}e^{-}$ background, we removed events for which the angular
separation in the calorimeter had no polar angle ($\theta$) component.

\begin{figure}[thb]
\centerline{\epsfig{file=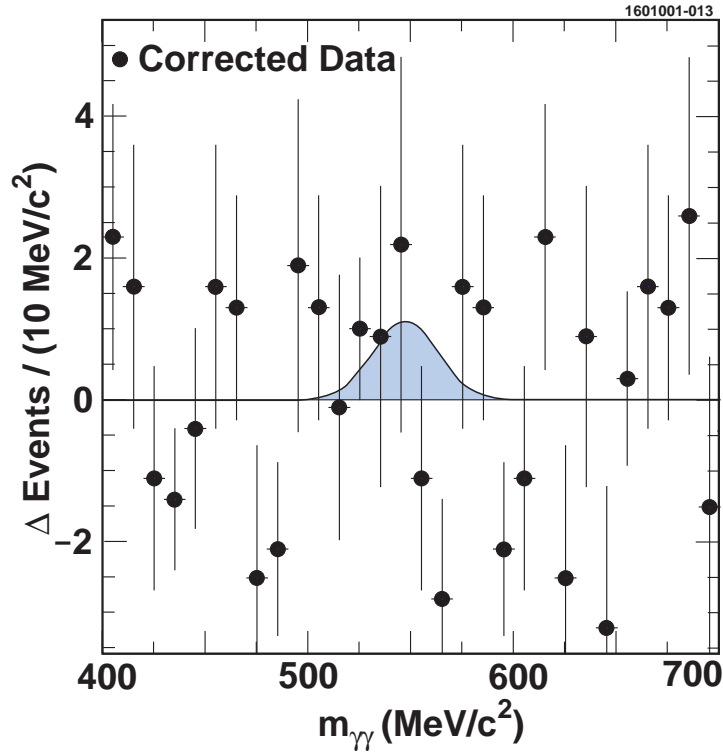, height=4.0in}}
\vspace{10pt}
\caption{Shown as solid circles is the  
di-photon spectrum for $\Upsilon$(1S) 
data after subtraction of the $\Upsilon$(4S) data, appropriately
scaled.
The $\Upsilon$(1S) data are fit, using a likelihood
technique, for a $\Upsilon$(1S)$\to\gamma\eta$ signal 
plus
the scaled $\Upsilon$(4S) spectrum, 
as detailed in the text.  
The superimposed curve shows the 
signal portion of the fit result.}
\label{subtraction_fit}
\end{figure}

To show that most of the remaining $\gamma\gamma\gamma$ background is of
QED 
or continuum
origin and not from the $\Upsilon$(1S), we took the higher-energy
data and performed the same analysis.  We then subtracted this 
spectrum of $m_{\gamma\gamma}$ from
the $\Upsilon$(1S) spectrum, after scaling it by  
the relative integrated luminosity, 
the relative reconstruction efficiencies,
and the relative production rates.  
The last of these is determined from 
$udsc$ continuum
simulations at the two energies, for which we counted the number of
events having one two-photon combination with an invariant mass within
$3\sigma_{\gamma\gamma}$ of $m_{\eta}$.  
The mass resolution of $\sigma_{\gamma\gamma}$ = 15.7 MeV/$c^{2}$ 
is taken from simulation of signal events.
The
result of this subtraction is shown in Figure \ref{subtraction_fit};
the integral of the entries in this figure is $4.3\pm 9.4$,
consistent with zero.

We then performed a binned likelihood fit of the $\Upsilon$(1S) 
data, using 
the value of the corresponding 
bin in the scaled distribution from the $\Upsilon$(4S) data
and a Gaussian signal function. 
Here the 
Gaussian had a mean of 
the established \cite{PDG2000} value
of the mass of the $\eta$ meson and a width taken from our simulations
of the signal ($\sigma_{\gamma\gamma} = 15.7$ MeV/$c^{2}$).
The result of the fit 
is shown in Figure~\ref{subtraction_fit}, yielding 
$4.0 \pm 3.8$ events. 
Bin-by-bin statistical variations in the  $\Upsilon$(4S) spectrum 
were taken into account by performing this procedure 
multiple times, each
time randomly assigning the number of events in each $\Upsilon$(4S)
bin,
according to the statistics of that bin. The 
distribution function used in determining the limit was
the average of these 
several
functions.

For this mode our overall
reconstruction efficiency from signal simulation, including
possible systematic biases and statistical uncertainties from the
simulation, is 
$\epsilon_{\gamma\gamma} = (27.7 \pm 1.9)\%.$ 

The major sources of possible systematic uncertainty
in our efficiency calculation for
$\Upsilon\to\gamma\eta$
are shown in Table \ref{systerr}.
These follow closely those detailed in 
our prior study of $\Upsilon\to\gamma\eta^{\prime}$\cite{etaprime}.
The 
degree of
uniformity and definition of the fiducial 
volume of the barrel calorimeter ($\pm$2.2\%) 
relates to our 
modeling the detector response to
the proper angular distribution for 
this
radiative $\Upsilon$ decay.
Uncertainties in 
charged-track reconstruction ($\pm$1.0\% per track),
and trigger effects ($\pm$2.0\%) 
were determined from previous
detailed CLEO studies 
of low-multiplicity $\tau$-pair and $\gamma\gamma$ events.
Similar studies allowed determination of the possible 
uncertainty in the reconstruction of $\pi^{0}$ and $\eta$ mesons from 
photons\cite{ProcarioBalest} ($\pm$3\% per meson);   
this relates to our ability to find and measure the
daughter photons and, in the case of the neutral
pions, to have the two-photon invariant mass be within the
$\pm 9\sigma_{\pi}$ window around the established $\pi^{0}$ mass.
Our ability to model the $E/p$ requirement in the 
$\gamma\pi^{+}\pi^{-}\pi^{0}$ 
final
state was assessed using charged pions from $K^{0}_{S}$ decays and 
assigned an uncertainty of $\pm 3.2\%$.
Shower leakage and other calorimeter effects make 
the mass distribution for $\Upsilon$ candidates asymmetric; based on
CLEO experience
with exclusive radiative $B$ meson decays\cite{Savinov} we have
assigned 
an uncertainty of $\pm 2\%$ regarding our ability to model these effects.
Based on a study of varying the mass resolution, $\sigma_{\gamma\gamma}$,
in fits to the data before all the final criteria were imposed, we
assign a systematic uncertainty of $\pm$5\% from this source.
These uncertainties were added in quadrature,
along with the statistical uncertainty associated with the size of
Monte Carlo samples, to obtain
the overall systematic uncertainty in the efficiencies. 

\begin{table}[hbt]
\caption{Systematic uncertainty contributions, as relative percentages,
to the 
efficiency for the studied decay modes for $\Upsilon\to\gamma\eta$. 
The combined uncertainties were obtained using quadrature addition.
}
\begin{center}
\begin{tabular}{|l|c|c|c|}
%\hline
Uncertainty source	& $\gamma\gamma$ & $\pi^{0}\pi^{0}\pi^{0}$ & $\pi^{+}\pi^{-}\pi^{0}$  	\\
\hline
Fiducial requirements	
& 2.2		& 2.2		& 2.2		\\
Track reconstruction	        
& ~-~		&~-~		& 2.0		\\
$\eta,\pi^{0}$ reconstruction from $\gamma\gamma$	
& 3.0		& 9.0		& 3.0		\\
$E/p$ criterion
&~-~		& ~-~		& 3.2		\\
Trigger	simulation
& 2.0		& 2.0		& 2.0		\\
$\Upsilon$ mass distribution 
& 2.0		& 2.0		& 2.0		\\
Variation of $\sigma_{\gamma\gamma}$ in the fit
& 5.0		&~-~		&~-~		\\
\hline
Monte Carlo statistics  
& 1.6		& 2.4		& 1.0		\\
\hline
\hline
Combined uncertainty		
& 7.0		& 10.0		& 6.0		\\
%\hline
\end{tabular}
\end{center}
\label{systerr}
\end{table}

\begin{figure}[thb]
%\centerline{\epsfxsize 4in \epsffile{ggsum.eps} }
\centerline{\epsfig{file=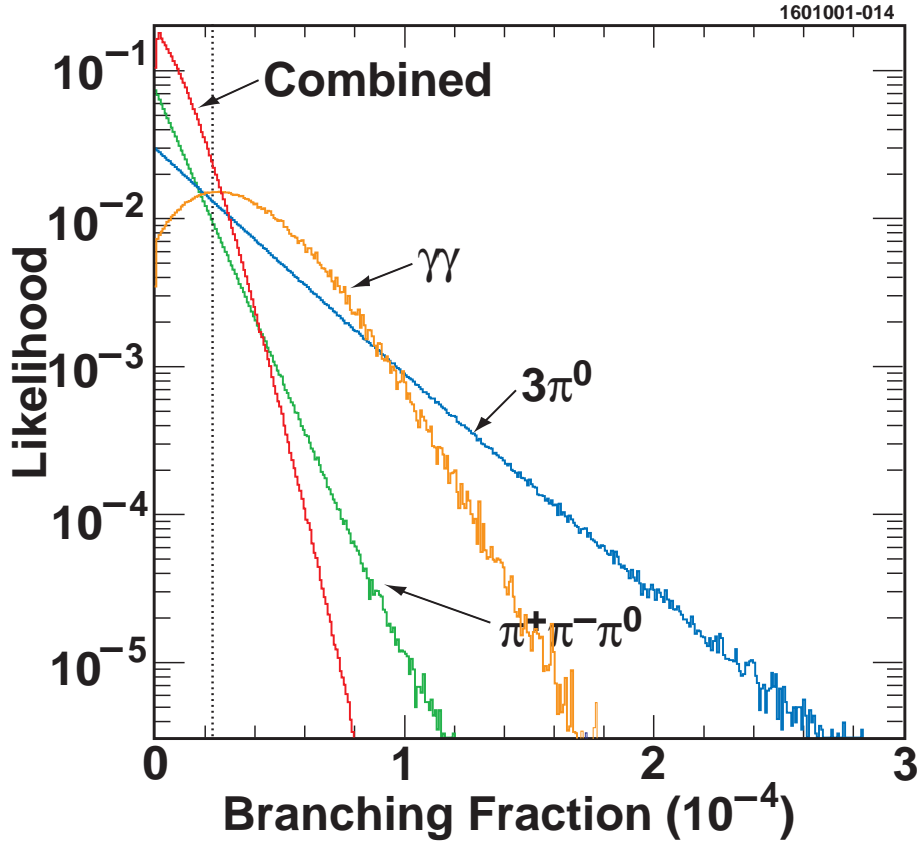, width=5.0in}}
%\centerline{\epsfxsize 4in \epsffile{1601001-014.eps} }
\vspace{10pt}
\caption{Likelihood functions for branching fraction from the three final 
modes studied in our analysis and the combined likelihood function. All 
distributions include smearing by systematic uncertainties and have been 
normalized to unit area.  The dotted vertical line is at 90\% of the
area of the combined function, namely $2.1\times 10^{-5}$.}
\label{likelihoods}
\end{figure}

The systematic uncertainties and the statistical uncertainty in the
number of $\Upsilon$(1S) decays are incorporated by a Monte Carlo
procedure to obtain likelihood distributions for the branching
fraction in each mode as 
${\cal B}(\Upsilon\to\gamma\eta) = N_{\eta}/(\epsilon N_{\Upsilon})$.  
In this approach we produce multiple experiments with $N_{\eta}$
from the likelihood function appropriate for
each decay mode\footnote{For the two $3(\pi)^{0}$ modes, which have no
events actually observed, this likelihood function is a falling exponential
as according to Poisson statistics.}
and then divide by an efficiency and by a number of $\Upsilon$(1S) events,
each picked from a Gaussian distribution about their mean values with
the appropriate standard deviations.  Summing the resulting likelihood
distributions, as shown in Figure~\ref{likelihoods}, to 90\% 
of their areas resulted in limits for 
$10^{5}\times{\cal B}(\Upsilon\to\gamma\eta)$ of
28.2, 6.7, 2.6, 1.9, and 2.1 for $\gamma\gamma\gamma$,
$\gamma\pi^{0}\pi^{0}\pi^{0}$,
$\gamma\pi^{+}\pi^{-}\pi^{0}$,
$\gamma (3 \pi)^{0}$,
and all combined, respectively.  The number of observed events,
detection efficiencies and limits are presented in Table~\ref{Eta_table}.

\begin{table}[hbt]
\caption{Results for the search of $\Upsilon\to\gamma\eta$. 
Results include statistical and systematic uncertainties, as
described in the text.}
\begin{center}
\begin{tabular}{|l|c|c|c|}

~	& $\gamma\gamma$ & $\pi^{0}\pi^{0}\pi^{0}$ & $\pi^{+}\pi^{-}\pi^{0}$ \\
\hline
observed events 	& 
$4.0\pm 3.8$ 		& 0	& 	0	\\
%\hline
${\cal B}_{\eta ,i}$ (\%)&  
$39.2\pm 0.3$ &$32.2\pm 0.4$ & $23.1\pm 0.5$  \\
Reconstruction efficiency (\%)&  
$27.7\pm 1.9$ & $7.6\pm 0.8$ & $26.7\pm 1.5$  \\
\hline
${\cal B}(\Upsilon\to\gamma\eta)$~(90\% C.L.) & 
$<28.2\times10^{-5}$	& $<6.7\times10^{-5}$ 	& $<2.6\times10^{-5}$	\\
\hline
Combined result	& \multicolumn{3}{c|}{$<2.1 \times10^{-5}$}	\\
\end{tabular}
\end{center}
\label{Eta_table}
\end{table}
%{\bf RSG to add a table here of efficiencies, limits, etc.}

\begin{figure}[thb]
\centerline{\epsfig{file=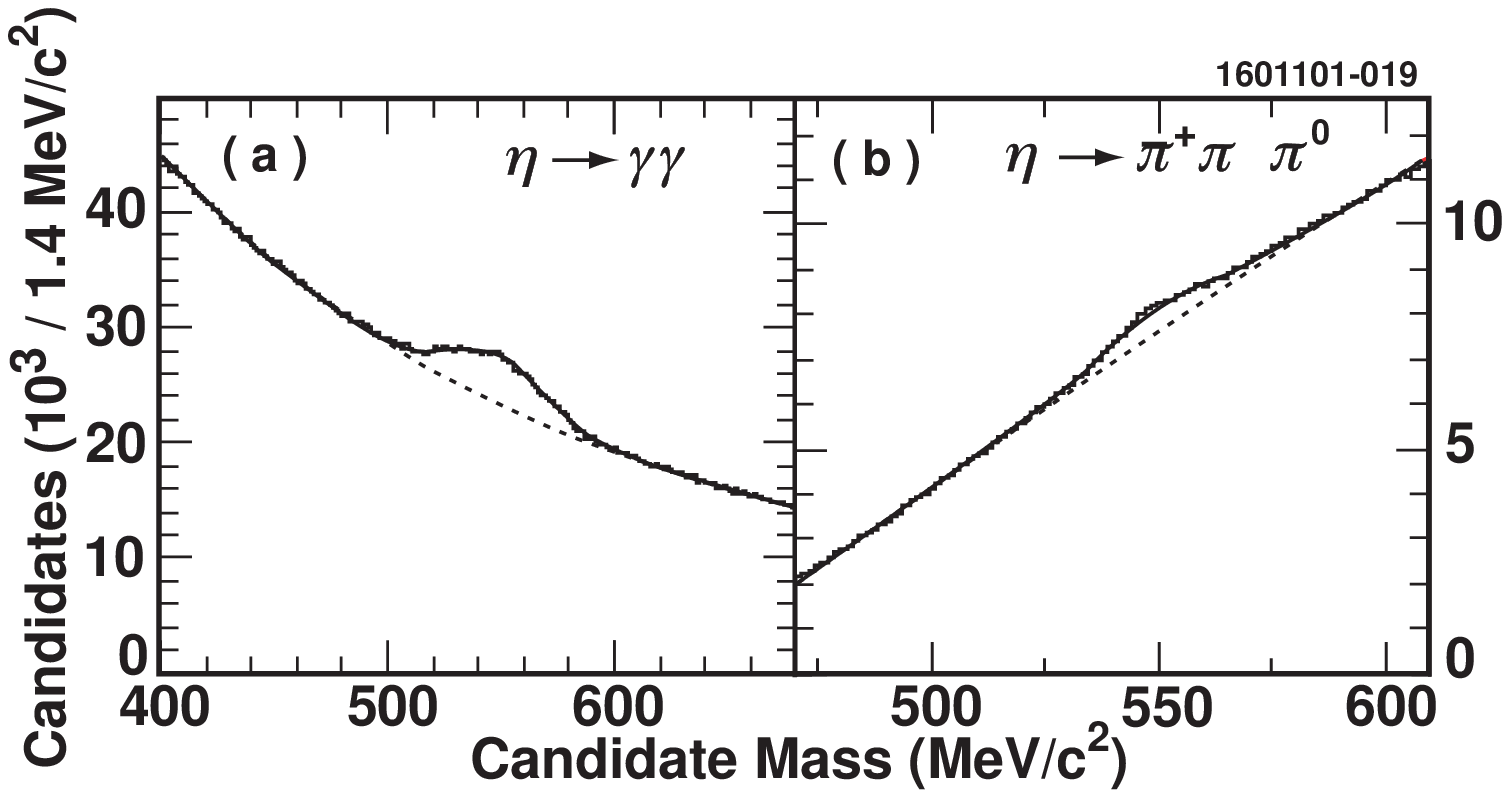, width=5.0in}}
%\centerline{\epsfig{file=1601001-015.eps, width=5.0in}}
%\centerline{\epsfxsize 3.5in \epsffile{1601001-015.eps} }
%\centerline{\epsfig{file=eta_etapr.ps,
%     height=4.0in}}
%    height=4.0in,
%    bbllx=80, bblly=300,bburx=550,bbury=750}}
\vspace{10pt}
\caption{The $\eta\to\gamma\gamma$ and 
$\eta\to\pi^{+}\pi^{-}\pi^{0}$ invariant mass distributions from
data taken at or near the $\Upsilon$(4S). The 
plots give the 
invariant mass distributions (histograms), which are each fit with 
the sum (solid lines) of a polynomial 
background (dashed lines) and a Gaussian signal.}
\label{etas}
\end{figure}

To show that we could use CLEO data to observe 
$\pi^{0}$ and $\eta$ mesons
we also applied our same selection criteria, with the exception of
requiring a high energy photon, to samples taken at the
$\Upsilon$(1S) and at or near the $\Upsilon$(4S).
Figure~\ref{etas} shows examples of inclusive yields of
$\eta$ mesons in the $\Upsilon$(4S) data that are consistent with
the expected rates\cite{PDG2000}. 

\section{Search for $\Upsilon\to\gamma$ {\protect f}$_{J}$(2220)}

Our search for 
$\Upsilon \to \gamma f_{J}$(2220) used
the three decay chains observed by BES\cite{BES}
that involve two charged tracks:
$f_{J} \to\pi^{+}\pi^{-}$,
$f_{J} \to K^{+}K^{-}$, and
$f_{J} \to p\overline{p}$.
We followed procedures very similar to those used
in our publication on the observation\cite{Korolkov} 
of two-body
radiative decays in
$\Upsilon \to \gamma \pi\pi$.

Events were required to have two quality tracks of 
opposite
charge with one energy deposition in excess
of 4 GeV in the barrel fiducial volume
($|\cos\theta|$ $\leq$ 0.71) of the detector. 
The events were also required to pass
trigger requirements\cite{CLEOTrig} 
that were highly efficient for this process and 
that could be reliably simulated.

\begin{figure}[thb]
\centerline{\epsfig{file=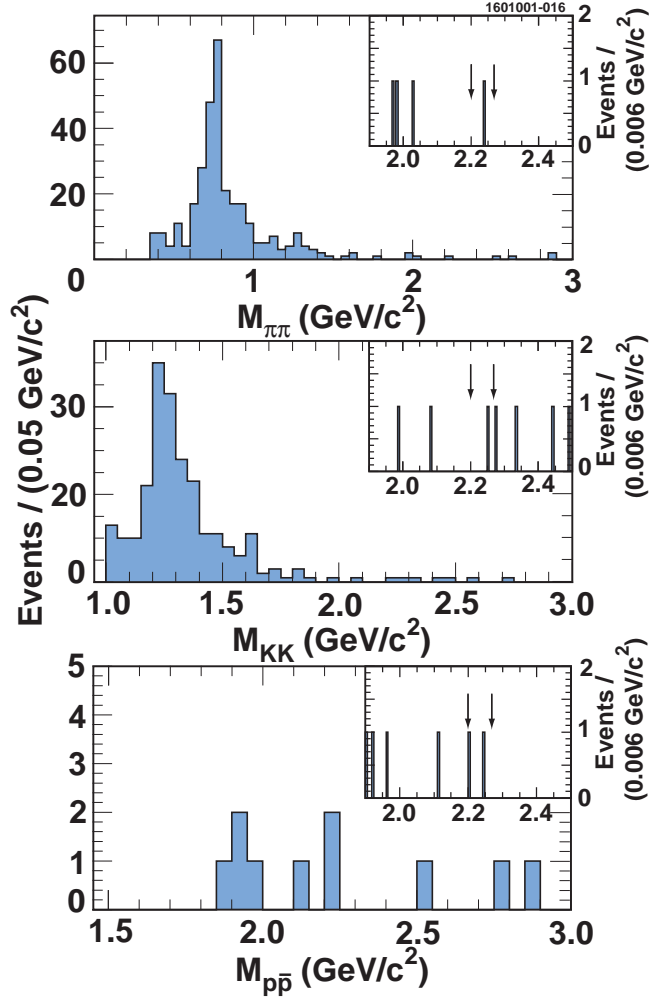, width=3.5in}}
%\centerline{\epsfxsize 3in \epsffile{1601001-016.eps} }
\vspace{10pt}
\caption{Di-hadron invariant mass distributions 
for data events passing all selection criteria. The inserts show
magnifications of the signal region. The vertical 
arrows in the inserts indicate the limits 
of the signal box, which are at $\pm2\Gamma$, $\Gamma = 17$~MeV/$c^2$.
No events are common to the signal regions of these channels.}
\label{pipikkpp}
\end{figure}

Backgrounds from QED processes such as 
$\gamma (\gamma)\mu^{+}\mu^{-}$ and $\gamma (\gamma) e^{+} e^{-}$
are potentially
large, so we next imposed criteria to minimize them.
At least one of the charged tracks had to extrapolate to the 
barrel muon detector and have momentum above 1.0 GeV/c and
yet {\it not} be identified as a muon.  For {\it each} track the
ratio of calorimeter energy to its measured momentum
had to satisfy $E/p < 0.85$ or $E/p > 1.05$, {\it i.e.,} inconsistent
with that expected for an electron.  The tracks could not be consistent
with coming from a photon conversion 
and had to have an opening
angle between them of less than 162$^{\circ}$.  For each of the 
three decay modes ($h = \pi^{+}, K^{+}, p$)  
we established four-momentum conservation
by demanding 
$-0.03 < (E_{\gamma h\overline{h}} - \sqrt{s})/\sqrt{s} < 0.02$
and $|\vec{p}_{\gamma h\overline{h}}| < 150$ MeV/$c$; these
represent approximately $\pm 3\sigma$ in our detector resolution and
take into account the effect of initial state radiation.
We did not 
explicitly
reject multiple combinations per event, and none
were observed in our simulations of the signal.  For each of the
three modes we then plotted the di-hadron invariant mass, as shown
in Figure~\ref{pipikkpp}.  
In that we will be comparing
our results to those of BES\cite{BES},  we use a 
mass of $m_{f} = $ 2.234~GeV/$c^2$ and a natural %dzb
width of 
$\Gamma_{f} = $17~MeV/$c^2$
for the $f_{J}$ resonance. Our ``signal box'' is defined as
$\pm 2 \Gamma_{f}$ centered at $m_{f}$.  
For comparison, our
signal simulations indicated mass resolutions ($\sigma$)
of $15.6 \pm 0.9$, $14.8 \pm 0.6$, and 
$9.9 \pm 0.4$ MeV/$c^{2}$ for the final states
$\pi^{+}\pi^{-}$, $K^{+}K^{-}$,  and $p\overline{p}$,
respectively.

The dominant continuum process ending in $\gamma\pi^{+}\pi^{-}$ is  
$e^+ e^- \to \gamma \rho$.
We used this reaction in two ways.  First, we 
repeated
the prior analysis
of two-body radiative decays\cite{Korolkov} to show that we could
reproduce the shape and magnitude of the $\gamma \rho$ enhancement.
To help us understand backgrounds we also studied 
data taken with an integrated luminosity of 838 pb$^{-1}$ at 
energies at or just below the $\Upsilon$(4S), recorded close in time 
to our $\Upsilon$(1S) data. 
Before subtracting, we 
scaled the normalization of the
$\pi^+\pi^-$ invariant mass distribution from 
the higher energy data by ${\cal L}/(s\epsilon)$ to take into account
luminosity, the energy 
dependence of the cross section, 
and the reconstruction efficiency. 
We observe an excess of ($29.4\pm 16.6$) events for the $\Upsilon$(1S)
data in 
0.45~GeV/$c^2<m_{\pi\pi}<$~1.1~GeV/$c^2$,
consistent with zero
excess.

We studied backgrounds to the 
$f_{J}$(2220) in three ways.  When we determined
upper limits we used the smallest of the three,
thereby being the most conservative.
In two of these, we used 
the data sets taken at or just below
the $\Upsilon$(4S) and accepted 
candidates within a background box width $\pm 10\Gamma_{f}$ 
centered at $m_{f}$. To understand the relative 
rates of production for background events at the two energies
($\sqrt{s}$ = 9.46 {\it vs.} 10.56 GeV), we first
simulated the $udsc$ continuum, including initial state
radiative effects, and for all events with a high-energy photon
counted the number in which the invariant mass recoiling against 
the photon was within $\pm 2 \Gamma_{f}$ for the $\Upsilon$(1S) energy, 
or within $\pm 10 \Gamma_{f}$ for the higher energy;\footnote{This 
is similar to the approach used in the study of
$\Upsilon\to\eta\gamma\to\gamma\gamma\gamma$ described in the
preceding section.} {\it i.e.,} we did not fix the photon energy but
rather the recoil mass region so that higher energy 
photons were involved in the
events at $\sqrt{s} = $10.56 GeV.
The ratio of these numbers is 0.17 as compared to the naive 
correction factor of 0.2 from the ratio of box sizes; {\it i.e.},
the relative production rate at the $\Upsilon$(1S) is 86\% of that
at the higher energy.  

If the dominant background in the signal region were from processes
such as $e^+ e^- \to\gamma\rho^{\prime}$ or $e^+ e^- \to\gamma\phi$, then
the scaling would be as $1/s$. In that case the
$\Upsilon$(1S) probability would be 124\% that
of the higher energy data.

There may also be 
background contributions from other, 
as yet unmeasured, radiative $\Upsilon$(1S) decays
({\it e.g.,} $\Upsilon\to \gamma f_{4}$(2050)), 
which are not accounted for when comparing
to the higher energy data. Therefore, as a third measure of
the background, we look in the sidebands of the $f_{J}$ region
of the di-hadron spectra from the $\Upsilon$(1S) data, namely
1.900~GeV/$c^2<m_{h\overline{h}}<$ 2.200~GeV/$c^2$ and
2.264~GeV/$c^2<m_{h\overline{h}}<$ 2.500~GeV/$c^2$. The number of events
found in this region is then scaled by the ratio of the bin widths
to predict the background in the signal region.
 
For the channel $f_{J} \to\pi^{+}\pi^{-}$, we observed one
candidate in the $\pm 2\Gamma_{f}$ signal region 
of the $\Upsilon$(1S) data set (see Figure~\ref{pipikkpp}a).
There were eight events in the broader $\pm 10\Gamma_{f}$ box at the
higher energies, which when scaled by luminosity, efficiency, and the
production ratio (as %dzb - described above
obtained from 
$udsc$ simulations, described above) gave a mean background level in the
signal box of $\mu_{b}(\pi\pi) = 0.12$ events. Using the $1/s$ scaling or
the sideband technique yielded 0.17 and 0.36 events, respectively.

Similarly for the channel $f_{J} \to\ K^{+} K^{-}$
we found one event in the signal box in the $\Upsilon$(1S)
data (see Figure~\ref{pipikkpp}b).  The analysis of the
higher energy data showed 14 events in the $\pm 10\Gamma_{f}$
background region, distributed uniformly.  When scaled appropriately
for luminosity, efficiency, and production rate (as described above),
this gave a mean background in the signal region of
$\mu_{b}(KK) = 0.21$ events.  
Using the $1/s$ scaling or
the sideband technique yielded 0.30 and 0.72 events, respectively.

The situation for $f_{J} \to\ p\overline{p}$ was  different.
Assigning proton masses to the two charged tracks moves the
large peak from $e^+ e^- \to \gamma \rho$ into the signal region.
Although our selection criteria for conservation of four-momenta
removed most of these backgrounds, we added a restriction that the
time-of-flight as measured in the scintillation system
be consistent for the two tracks to have proton masses, given their
measured momenta.  Also, nucleon-antinucleon annihilation can deposit 
large energies in
the electromagnetic calorimeter, so our requirement on $E/p$ would be
very inefficient for anti-protons; therefore this requirement was removed
for the negative track.
The resulting distribution of $p\overline{p}$ invariant mass is shown in 
Figure~\ref{pipikkpp}c, having two candidate events in the signal box
region of $\pm 2\Gamma_{f}$.  In the higher energy data the background
was no longer linear, due largely
to the feed-through of $e^+ e^- \to \gamma \rho$
events.  We therefore fit this distribution to a Gaussian for this
$\gamma\rho$ portion and a flat contribution.  Integrating the fit
result over the signal box region and correcting 
for relative efficiencies, integrated luminosities, 
and production (as described above)
gave a mean background for this mode of 
$\mu_{b}(p\overline{p}) = 0.28$ events.  A $1/s$ scaling
would imply 0.40 events of background in the signal region,
while the sideband technique
gave a somewhat larger result of 0.48 events.
	
In analyzing these three decay modes, we found that the 
signal regions are kinematically distinct so that no events
are common to any two of them.

To calculate reconstruction efficiencies we used our Monte Carlo
simulations, based again on GEANT\cite{GEANT},
with $\Upsilon\to\gamma f_{J}$
and 
$f_{J} \to h\overline{h}$ for each
of $h = \pi^{+}, K^{+}, p$. The efficiencies, including possible systematic
biases and uncertainties
(as discussed below and summarized in Table~\ref{fjsyst})
and the statistics of the Monte Carlo
samples, were 
(28.8 $\pm $4.2)\%,
(21.9 $\pm $3.8)\%, and
(27.2 $\pm $3.3)\%, respectively.

The simulation of signal events 
was
generated uniformly 
in $\cos \theta$ of the high-energy photon. 
Using the analysis for
$J/\psi \to\gamma f_{J}$ with $J=2$\cite{Einsweiler}
we have evaluated the range of possible angular distributions
and their effect on our calculation of the geometric
acceptance, assigning a systematic uncertainty of $\pm10$\%. 
As in our search for $\Upsilon\to\gamma\eta$, we use previous
detailed studies of low multiplicity $\tau^{+}\tau^{-}$ and 
$\gamma\gamma$ events to study uncertainties in track reconstruction
and momentum measurement
($\pm 1.0\%$ per track)
and trigger simulation ($\sim \pm 2.6\%$,
and slightly dependent on momentum of the charged daughters). Also,
as before, we use previous %dzb -published 
CLEO experience\cite{Savinov} with radiative $B$
meson decays to assess possible biases ($\pm 2\%$) in our 
$\Upsilon$ invariant mass determination due to shower leakage and
other calorimeter effects. 

To evaluate the correctness of our simulations of the electron
and muon rejection criteria, we
analyzed charged pions from 
$\tau^{-}\to\rho^{-}\nu_{\tau}$ with 
$\rho^{-}\to\pi^{-}\pi^0$ (and charged conjugates)
for both data and Monte Carlo.
Using restrictive selection criteria on the masses of the 
$\pi^{0}$ and $\rho^{-}$ mesons, on the specific ionization ($dE/dx$) of the
charged pion, and the opening angle between the pions, we are left
with a sample of charged tracks which is, according to the simulation, 
over 98\% pions with less than 1\% each of electrons and muons.
Weighting the observed 
differences between simulation and data for these $\tau$ data by the
momentum spectra of the pions in the $f_{J}$ simulation shows that
we need to increase our efficiencies by 2.6\% and 9.9\% due to the
effects of the $E/p$ and muon requirements, respectively.  
We also assigned a systematic
uncertainty of that same magnitude for these possible biases.
For the final states $K^{+}K^{-}$ and $p\overline{p}$,
we scaled these uncertainties by the relative inefficiencies
for events to pass these two criteria in our signal
simulation, as indicated in Table~\ref{fjsyst}.
We note that the magnitude of the uncertainty stemming from
the muon rejection criteria is similar to that from our prior
work\cite{Korolkov} on $\Upsilon\to\gamma\pi^{+}\pi^{-}$, which was $\pm$13\%,
and calculated in an independent fashion.
 
To estimate the systematic uncertainties associated with possible 
mismodeling of the time-of-flight identification
we followed the strategy used in our 
study\cite{Ong}
of $\gamma\gamma\to p\overline{p}$. Here we varied the 
widths of the timing distributions by $\pm 20$\%, which is about
four times the precision to which they are known. This changes 
the reconstruction efficiency by $\pm2$\%, which we assigned as the
systematic uncertainty from this source.

\begin{table}[htb]
\caption{Systematic uncertainty contributions, as relative percentages,
to the 
efficiency for the studied decay modes for $\Upsilon\to\gamma f_{J}$(2220). 
The combined uncertainties were obtained using quadrature addition.
}
\begin{center}
\begin{tabular}{|l|ccc|}
Error source	& 
$f_{J}\to\pi^{+}\pi^{-}$	& 
$f_{J}\to\ K^{+} K^{-}$		& 
$f_{J}\to\ p\overline{p}$	\\
\hline%-----------------------------------------------------------------------------------
Angular distribution		& 10.0\%	& 10.0\%	& 10.0\%\\
Trigger simulation		& 2.6\%		& 2.6\%		& 2.7\%	\\
Track reconstruction		& 2.0\%		& 2.0\%		& 2.0\%	\\
$\Upsilon$ mass distribution	& 2.0\%		& 2.0\%		& 2.0\%	\\
$E/p$ criterion			& 2.6\% 	& 1.7\%		& 0.3\%	\\
Muon suppression		& 9.9\%		& 13.8\% 	& 5.7\%	\\
TOF identification		& --		& --		& 2.0\%	\\
\hline%-----------------------------------------------------------------------------------
Monte Carlo statistics 		& 1.0\%		& 1.2\%		& 0.9\%	\\			
\hline%-----------------------------------------------------------------------------------
Total (quadrature sum)		& 14.8\%	& 17.6\%	& 12.3\%\\
\end{tabular} 
\end{center}
\label{fjsyst}
\end{table}

To determine confidence limits for the results of this 
analysis, we followed
the method advocated by Feldman and Cousins~\cite{Cousins}, which 
avoids under-coverage of confidence intervals. 
We adapted 
their method by replacing the confidence intervals for different 
mean numbers of observed events, $\mu$,  
by confidence intervals for different 
values of the branching fraction,
${\cal B}=\mu /(N_{\Upsilon}\epsilon)$.  
The uncertainties in the efficiencies and in the 
number of $\Upsilon$(1S)  produced were incorporated
by smearing the central values with Gaussian distributions. Finally, 
we extend the confidence limits so that they cover the integer values 
allowed by Poisson statistics.

Before including systematic effects, we obtain 
90\% confidence intervals for the product
${\cal B}(\Upsilon\to\gamma f_{J})
\times
{\cal B}(f_{J}\to h\overline{h})$ of 
${\cal B}{\cal B}<10.2\times10^{-6}$ for $\pi^{+}\pi^{-}$,
${\cal B}{\cal B}<13.1\times10^{-6}$ for $K^{+} K^{-}$, 
and the range 
$0.7\times10^{-6}<{\cal B}{\cal B}<14.3\times10^{-6}$ 
for $p\overline{p}$.  
Note that for the
$p\overline{p}$ case we have an interval with both lower and upper
limits. After adding these systematic effects, we obtain
${\cal B}{\cal B} <12.0\times10^{-6}$ for $\pi^{+}\pi^{-}$,
${\cal B}{\cal B} <15.5\times10^{-6}$ for $K^{+} K^{-}$, 
and the range 
$0.5\times10^{-6}<{\cal B}{\cal B}<16.2\times10^{-6}$ 
for $p\overline{p}$.
As noted earlier, we have used the estimate of the backgrounds,
$\mu_{b}$, as determined by the simulated production ratios.
Using larger background estimates, such as from $1/s$ scaling or
sideband comparisons, would give lower upper limits and eliminate
the lower limit in the case of $f_{J}\to p\overline{p}$;  
for these three background estimates the probability of observing
two or more events is between 3 and 8\%, insufficiently small to
claim a signal in this decay mode.
We 
therefore take the conservative approach and quote the 90\%
confidence level upper limits as given and quote no lower limit
for any of the channels.
The results of our analysis are summarized in Table~\ref{resultstab}.

\begin{table}[htb]
\caption{Results of the search for $\Upsilon\to\gamma f_{J}$.
The estimated background, obtained by scaling by the simulated
continuum production rates, is the smallest of the
three estimates of background levels and leads to the most
conservative upper limits.The efficiencies and limits include 
systematic effects.  
The last row of entries is a scaling of the BES results
for $J/\psi$ decays.}
\vspace{-12pt}
\begin{center}
\begin{tabular}{|l|ccc|}			
~& 
$f_{J}\to\pi^{+}\pi^{-}$ & 
$f_{J}\to\ K^{+} K^{-}$ & 
$f_{J}\to\ p\overline{p}$			\\
\hline
Observed events in $\pm 2\Gamma_{f}$	& 1	& 1	& 2 \\
Scaled continuum background ($\mu_{b}$)	        & 0.12	& 0.21	& 0.28	\\
Overall efficiency & ($28.8\pm4.2$)\% & ($21.9\pm3.8$)\% &($27.2\pm3.3$)\% \\
${\cal B}(\Upsilon\to\gamma f_{J}$)$\times$
& ~& ~ & ~ \\
~~~~~${\cal B}(f_J(2220) \to h\overline{h}$) (90\%~C.L.)
& $<12.0\times10^{-6}$	& $<15.5\times10^{-6}$	& $<16.2\times10^{-6}$	\\
\hline
${\cal B}(J/\psi ~\to\gamma f_{J}) \times$ &
~& ~ & ~ \\
~~~~${\cal B}(f_J(2220) \to h\overline{h})\times 0.04$& 
$1.5^{+0.9}_{-0.8}\times 10^{-6}$    &
$2.5^{+1.2}_{-1.1}\times 10^{-6}$    &
$0.7\pm 0.3\times 10^{-6}$     \\
\end{tabular}
\end{center}
\label{resultstab}
\end{table}

As a comparison we have also computed the upper limits following the 
traditional approach of summing up the Poisson distribution 
${\cal P}(\leq n|~\mu) = e^{-\mu}\times 
(1 + \mu + ... + \frac{\mu^{n}}{n!})$ to 
90\% C.L. 
For $n=1$ and $2$, this gives limits of $\mu=3.88$ and $5.32$, 
respectively. 
Ignoring backgrounds the corresponding upper limits on the product
branching fractions, including systematic effects,
are
${\cal B}{\cal B}<9.3\times10^{-6}$ for $\pi^{+}\pi^{-}$, 
${\cal B}{\cal B}<12.2\times10^{-6}$ for  $K^{+} K^{-}$, and
${\cal B}{\cal B}<13.5\times10^{-6}$ for $p\overline{p}$. 
Note that
this method can lead to 
under-coverage, as outlined in the paper by Feldman and 
Cousins~\cite{Cousins}.

\section{Comparison to Prior Results and Theory}

The only other reported analysis~\cite{PDG2000} of 
${\cal B}(\Upsilon\to\gamma\eta)$
is by
Crystal Ball~\cite{xbal}, which determined a 90\% C.L. upper limit of 
$3.5 \times 10^{-4}$; our limit of $2.1 \times 10^{-5}$ is $\sim 17$ 
times more 
stringent.

We can then use our new limit on ${\cal B}(\Upsilon\to\gamma\eta)$,
the 
measured enhancement\cite{Korolkov} near 1270 MeV/$c^{2}$ in
$\Upsilon\to\gamma\pi\pi$, and the measurements 
of
${\cal B}$($J/\psi \to \gamma f_{2}$(1270))$ = (1.38 \pm 0.14) \times 10^{-3}$ 
and
${\cal B}$($J/\psi \rightarrow \gamma\eta) = (0.86 \pm 0.08) \times 10^{-3}$~\cite{PDG2000}
to create interesting ratios.
Here we assume that the observed structure in the di-pion mass
spectrum near 1270 MeV/$c^{2}$ in the $\Upsilon$ 
study is totally due to $f_{2}$
production, which implies
${\cal B}$($\Upsilon\to\gamma f_{2}$(1270))$ = (8.2 \pm 3.6) \times 10^{-5}$.
We then obtain
\begin{equation}
R_{f}(\Upsilon,J/\psi) \equiv \frac{{\cal B}(\Upsilon \to \gamma f_{2})} {{\cal B}(J/\psi \to\gamma f_{2})}
= 0.061 \pm 0.026~~~\mathrm{and}
\end{equation}

\begin{equation}
R_{\eta}(\Upsilon,J/\psi) \equiv \frac{{\cal B}(\Upsilon \to \gamma\eta)} {{\cal B}(J/\psi \to\gamma\eta)}
< 0.024~(90\%~\mathrm{C.L.});
\end{equation}

\noindent and, looking instead at the ratios of the two final states
for a given vector parent, 

\begin{equation}
R_{\Upsilon}(\eta,f_{2}) \equiv \frac{{\cal B}(\Upsilon\to\gamma\eta)} {{\cal B}(\Upsilon\to\gamma f_{2})}
< 0.32~(90\%~\mathrm{C.L.})~~~\mathrm{and}
\end{equation}

\begin{equation}
R_{J/\psi}(\eta,f_{2}) \equiv \frac{{\cal B}(J/\psi \to \gamma\eta)} {{\cal B}(J/\psi \to \gamma f_{2})} 
= 0.62 \pm 0.09.
\end{equation}

\noindent These results were obtained by including the various
uncertainties in a Monte Carlo technique.
In the last two of the ratios, we assume 
that all the uncertainties from the $J/\psi$ and $\Upsilon$ measurements 
are uncorrelated. 
Our limits show that the branching fractions into $\gamma\eta$ and
$\gamma f_{2}$(1270) behave differently in the 
cases of $J/\psi$ and $\Upsilon$,
although not as dramatically as in the case of 
$\Upsilon\to\gamma\eta^{\prime}$\cite{etaprime}.
In addition we form the double ratio:

\begin{equation}
{\cal R} \equiv
\frac{R_{\Upsilon}(\eta, f_{2})}{R_{J/\psi}(\eta, f_{2})}
= \frac
{{\cal B}(\Upsilon\to\gamma\eta)}
{{\cal B}(\Upsilon\to\gamma f_{2})}
\times
\frac
{{\cal B}(J/\psi \to \gamma f_{2})}
{{\cal B}(J/\psi \to \gamma\eta)}
< 0.53
\end{equation}

\noindent at 90\% C.L.  This is 
to be compared
with the prediction of K\"orner {\it et al.}~\cite{KKKS} of

\begin{equation}
{\cal R}_{theory} = \frac{0.10}{0.24} = 0.42 ;
\end{equation}

Chao's technique~\cite{Chao} first calculates mixing angles among the
various pseudoscalars, extending the $J^{PC} = 0^{-+}$ nonet
to include heavier cousins such as the $\eta_{b}$.  
Then, using the predicted allowed M1 transition $\Upsilon\to\gamma\eta_{b}$,
Chao predicts ${\cal B}(\Upsilon\to\gamma\eta) = 1 \times 10^{-5}$,
which is consistent with our limit.  We note that in our
prior work~\cite{etaprime} we did not know of Chao's 
prediction of ${\cal B}(\Upsilon\to\gamma\eta^{\prime}) = 6 \times 10^{-5}$,
which is 
to be compared 
with our upper limit for that process
of $1.6 \times 10^{-5}$ at 90\%~C.L.

Intemann's extended vector dominance model gives
$6.5 \times 10^{-8} 
< {\cal B}(\Upsilon\to\gamma\eta) 
< 1.2 \times 10^{-7}$, with the two limits determined by
having destructive or constructive interference, respectively,
between the
terms involving $\Upsilon$ and $\Upsilon^{\prime}$.  This range
of predictions is well below our new limit.
Ma uses a technique in which the decay amplitude factorizes into a
non-relativistic piece describing the bound state of the heavy
quarkonium and an expansion in ``twist'' to characterize the 
conversion of the gluons into the final state meson.  The 
published value~\cite{Ma} 
was ${\cal B}(\Upsilon\to\gamma\eta) = 1.2 \times 10^{-7}$,
although subsequent correspondence~\cite{Ma2} indicates 
the correct value is actually
$3.3 \times 10^{-7}$.  In either case, these are significantly below our
limit.
Feldmann {\it et al.}\cite{FKS} predict the ratio
$\Gamma (\Upsilon\to\gamma\eta^{\prime})/\Gamma (\Upsilon\to\gamma\eta)$
= 6.5.  Given that we only have limits on these two processes, we cannot
address their prediction.

For the glueball candidate,
our 90\% C.L. limits on the product branching fractions
${\cal B}(\Upsilon\to\gamma f_{J})
\times
{\cal B}(f_{J}\to h\overline{h})$
are on the order of $1.5\times 10^{-5}$ for each of the three
modes.  In Table~\ref{resultstab} we show the results from BES\cite{BES}
for these channels in radiative $J/\psi$ decay, scaled by a factor
of ${\cal F}$.  This scaling 
arises from the naive expectation that the
amplitude for the radiative process of meson formation varies directly
as the quark charge (to couple to the photon) and inversely as the 
quark mass (from the fermionic propagator between the photon emission
and the resonance formation).  One then squares this to get the 
rate and corrects for the full widths of the heavy 
quarkonia\cite{PDG2000} to obtain 

\begin{equation}
{\cal F} = (\frac{q_{b}m_{c}}{q_{c}m_{b}})^{2}\cdot
\frac{\Gamma_{J/\psi}}{\Gamma_{\Upsilon}} = 0.04 .
\end{equation}

Here we have used masses of 1.7 GeV/$c^{2}$
and 5.2 GeV/$c^{2}$ for the charm and bottom quarks,
respectively.  As shown in Table \ref{resultstab}
and in Figure \ref{SummaryFigure}, our limits on radiative
$f_{J}$(2220) production do not confront these
predictions.

As a quantification of this situation, we have have used a Monte Carlo
technique to evaluate the ratio of the CLEO and BES\cite{BES} results, 
{\it i.e.},
 
\begin{equation}
R_{f_{J}}(\Upsilon, J/\psi) \equiv
\frac{{\cal B}(\Upsilon\to\gamma f_{J})}{{\cal B}(J/\psi~\to\gamma f_{J})}
=\frac
{{\cal B}(\Upsilon\to\gamma f_{J})\cdot{\cal B}(f_{J}\to h\overline{h})}
{{\cal B}(J/\psi~\to\gamma f_{J})\cdot{\cal B}(f_{J}\to h\overline{h})} .
\end{equation}

For the denominator we add the BES statistical and systematic uncertainties
in quadrature and throw BES ``experiments'' in a Gaussian distribution,
keeping only physical values.  In all three cases ($h = \pi^{+}, K^{+}, p$)
we can only say that $R_{f_{J}}$ is less than roughly 0.50
at 90\% C.L., an order of magnitude
from the predicted ratio of ${\cal F} = 0.04$.

We also note that the Crystal Barrel\cite{XBarrel} collaboration has
combined their results with those of BES\cite{BES} to obtain that
${\cal B}(J/\psi\to\gamma f_{J}$(2220)$) > 0.003$ at 95\% C.L.  Applying
the naive scaling factor ${\cal F}$ would then predict that
${\cal B}(\Upsilon\to\gamma f_{J}$(2220)$) > 1.2\times 10^{-4}$; this
would be larger than the CLEO result\cite{Korolkov} 
for the radiative decay to $f_{2}$(1270).

\section{Summary and Acknowledgments}

In summary, we have used the CLEO detector operating at the CESR
storage ring to search for two-body radiative $\Upsilon$(1S)
decays.  In the work presented in this Article, we reported
specifically on searches for the decay
$\Upsilon\to\gamma\eta$ with the subsequent decays
$\eta\to\pi^{0}\pi^{0}\pi^{0}$,
$\eta\to\pi^{+}\pi^{-}\pi^{0}$, and
$\eta\to\gamma\gamma$ and for the decay 
$\Upsilon\to\gamma f_{J}$(2220) with the subsequent decays
of the glueball candidate of
$f_{J}\to\pi^{+}\pi^{-}$,
$f_{J}\to K^{+} K^{-}$, and
$f_{J}\to p\overline{p}$.
Including our prior published results we have the following:

for the decay $\Upsilon\to\gamma\pi\pi$\cite{Korolkov},
$$
{\cal B}(\Upsilon\to\gamma\pi^{+}\pi^{-}) = 
(6.3\pm 1.2 \pm 1.3) \times 10^{-5}~~[m_{\pi\pi} > 1.0~{\rm GeV/}c^{2}]~,
$$
and 
$$
{\cal B}(\Upsilon\to\gamma\pi^{0}\pi^{0}) = 
(1.7\pm 0.6 \pm 0.3) \times 10^{-5}~~[m_{\pi\pi} > 1.0~{\rm GeV/}c^{2}]~;
$$
and, assuming the enhancement in the 
invariant mass spectrum for $\pi^{+}\pi^{-}$ 
in the region of 1.3 GeV/$c^{2}$ is due to 
$f_{2}$(1270) production,

$$
{\cal B}(\Upsilon\to\gamma f_{2}(1270)) 
= (8.2 \pm 3.6) \times 10^{-5} ;
$$
for the decay  $\Upsilon\to\gamma\eta^{\prime}$\cite{etaprime},
$$
{\cal B}(\Upsilon\to\gamma\eta^{\prime}) < 1.6 \times 10^{-5}
~~[90\% {\rm C.L.}];
$$
for the decay  $\Upsilon\to\gamma\eta$,
$$
{\cal B}(\Upsilon\to\gamma\eta) < 2.1 \times 10^{-5}
~~[90\% {\rm C.L.}];
$$
and for the product branching fractions involving the
glueball candidate $f_{J}$(2220),
$$
{\cal B}(\Upsilon\to\gamma f_{J})
\times
{\cal B}(f_{J}\to\pi^{+}\pi^{-}) < 1.2 \times 10^{-5}~~ [90\% {\rm C.L.}],
$$
$$
{\cal B}(\Upsilon\to\gamma f_{J})
\times
{\cal B}(f_{J}\to K^{+} K^{-}) < 1.6 \times 10^{-5}~~ [90\% {\rm C.L.}],
$$
and
$$
{\cal B}(\Upsilon\to\gamma f_{J})
\times
{\cal B}(f_{J}\to p\overline{p})< 1.6 \times 10^{-5}~~ [90\% {\rm C.L.}].
$$

\begin{figure}[thb]
\centerline{\epsfig{file=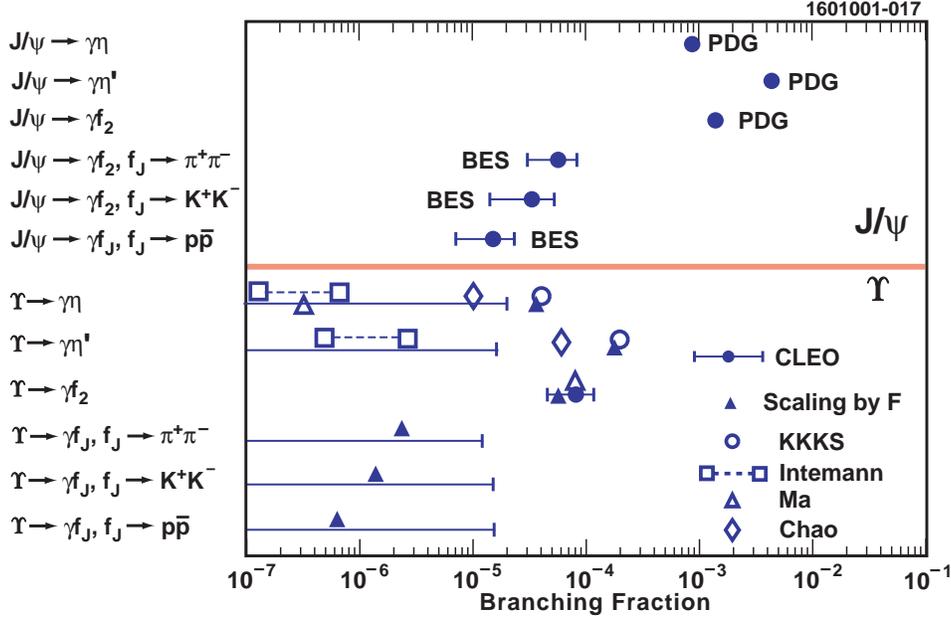, width=5.0in}}
\vspace{10pt}
\caption{Radiative decays of $J/\psi$ and $\Upsilon$(1S) vector mesons.
Shown are the experimental results (as solid circles or limit
bars)
from the PDG for the 
well-established radiative $J/\psi$ decays to $\eta$, $\eta{\prime}$,
and $f_{2}$(1270), from BES for the three charged modes of
radiative decay to the glueball candidate $f_{J}$(2220),
and from CLEO for the $\Upsilon$ decays.  
The solid triangles give the values for
radiative $\Upsilon$ decay based on radiative $J/\psi$ decay 
and the naive scaling involving the masses and charges of the
constituent quarks.  Various explicit theoretical predictions
for the radiative production of $\eta$, $\eta^{\prime}$, and
$f_{2}$(1270)
are shown as open symbols.
For references see the text.}
\label{SummaryFigure}
\end{figure}

We also present this summary in graphical form in Figure~\ref{SummaryFigure}:
({\it i}) the world-average\cite{PDG2000} values for the radiative 
two-body decays of the $J/\psi$ to $\eta$, $\eta^{\prime}$, and $f_{2}$(1270);
({\it ii}) the BES\cite{BES} results for the product branching fractions
${\cal B}(J/\psi\to\gamma f_{J})\times{\cal B}(f_{J}\to h^{+}h^{-})$;
({\it iii}) the scaling of these $J/\psi$ results by ${\cal F}$ to obtain
estimates for the corresponding $\Upsilon$(1S) decays; 
({\it iv}) the CLEO results 
%just 
summarized above; and 
({\it v}) theoretical predictions from  K\"orner {\it et al.}\cite{KKKS},
Intemann\cite{Intemann}, Ma\cite{Ma}, and Chao\cite{Chao}.

We gratefully acknowledge the effort of the CESR staff in providing us with
excellent luminosity and running conditions.
M. Selen thanks the PFF program of the NSF and 
the Research Corporation, 
and A.H. Mahmood thanks the Texas Advanced Research Program.
This work was supported by the National Science Foundation, the
U.S. Department of Energy, and the Natural Sciences and Engineering Research 
Council of Canada.

\end{document}